# Observational and Theoretical Studies of 27 δ Scuti Stars with Investigation of the Period-Luminosity Relation

**Atila Poro[1,2,3], Ehsan Paki[1,3], Golnaz Mazhari[1,3], Soroush Sarabi[1,2], Filiz Kahraman Alicavus[4,5], Farzaneh Ahangarani Farahani[1,3], Hamidreza Guilani[3], Alexander A. Popov[6], Alexandra M. Zubareva[7,8], Behjat Zarei Jalalabadi[1,3], Mahshid Nourmohammad[3], Fatemeh Davoudi[1,2,3], Zahra Sabaghpour Arani[3], Amir Ghalee[3,9]**

[1]The International Occultation Timing Association Middle East section, Iran, info@iota-me.com
[2]Astronomy Department of the Raderon Lab., Burnaby, BC, Canada
[3]The Eighth IOTA/ME Summer School of Astronomy, Tafresh University, Tafresh, Iran
[4]Çanakkale Onsekiz Mart University, Faculty of Sciences and Arts, Physics Department, 17100, Çanakkale, Turkey
[5]Çanakkale Onsekiz Mart University, Astrophysics Research Center and Ulupınar Observatory, TR-17100, Çanakkale, Turkey
[6]Kourovka Astronomical Observatory of Ural Federal University, Ekaterinburg, Russia
[7]Institute of Astronomy, Russian Academy of Sciences, Moscow, Russia
[8]Sternberg Astronomical Institute, Lomonosov Moscow State University, Moscow, Russia
[9]Department of Physics, Tafresh University, P.O. Box 39518-79611, Tafresh, Iran

**Abstract**
The multi-color CCD photometric study of 27 δ Scuti stars is presented. By using approximately three years of photometric observations, we obtained the times of maxima and magnitude changes during the observation time interval for each star. The ephemerides of our δ Scuti stars were calculated based on the Markov Chain Monte Carlo (MCMC) method using the observed times of maxima and the period of the stars' oscillations. We used the Gaia EDR3 parallaxes to calculate the luminosities and also the absolute magnitudes of these δ Scuti stars. The fundamental physical parameters of all the stars in our sample such as masses and radii were estimated. We determined the pulsation modes of the stars based on the pulsation constants. Moreover, the period-luminosity ($P-L$) relation of δ Scuti stars was investigated and discussed. Then, by using a machine learning classification, new $P-L$ relations for fundamental and overtone modes are presented.

## 1. Introduction

Delta Scuti stars are a type of pulsating variable stars with spectral types ranging from A0 to F5. They are primarily located in the lower part of the Cepheids instability strip in the Hertzsprung-Russell (H-R) diagram (Rodríguez & López-González 2000). δ Scuti stars are short-period variables that show pulsations generally in a range of 0.02 to 0.25 days with the pulsating amplitude ranging from 0.003 to 0.9 magnitude in the $V$-band (McNamara et al. 2000). δ Scuti stars have an effective temperature range of ~6300-8600 K, and they can be in different evolutionary stages with masses varying from ~1.6 $M_\odot$ for shorter period stars to ~2.4 $M_\odot$ for longer period stars (McNamara 2011).

Dwarf or subgiant classical δ Scuti stars are known as population I stars. Besides, SX Phe stars are the Population II stars which generally consist of evolved stars found in globular clusters (Pena et al. 1999). These stars are located in the lower part of the δ Scuti instability strip mixed with the δ Scuti stars. According to the pulsation amplitude, δ Scuti stars are divided into two subgroups; Low Amplitude Delta Scuti stars (LADS), and High Amplitude Delta Scuti stars (HADS). HADS pulsations have amplitudes of greater than 0.3 magnitude, whereas LADS pulsations have amplitudes of less than 0.1 magnitude in the $V$-band (Rodríguez et al. 1996). HADS are often subgiant stars while LADS can be on or close to the Main Sequence (MS) in the H-R diagram. The majority of δ Scuti stars show radial and/or non-radial low-order pressure ($p$) mode pulsations (Joshi & Joshi 2015).

In the lower part of the classical Cepheid instability strip, studying the changes in the pulsation period of δ Scuti stars helps to understand the stellar evolution (Breger & Pamyatnykh 1998). The first $P-L$ relation for Cepheid variables was published by Leavitt and Pickering (1912); continuing the study, different scientists have presented empirical $P-L$ relations like Fernie (1992), Laney (2000), and McNamara (1997, 1999, 2007, 2011). They have derived several $P-L$ relations by combining Cepheids with δ Scuti stars. Recent studies, Ziaali et al. (2019) and Jayasinghe et al. (2020) have obtained newly updated $P-L$ relations for these δ Scuti stars using the Gaia DR2 parallaxes.

In this study, we present a multi-color photometric analysis for 27 δ Scuti stars that have been found by observations performed in Russia and the USA from 2012 to 2015 (Popov et al. 2017). Burdanov et al. (2016) have examined all these observed stars to search for variable stars near the instability strip by using the Robust Median Statistics (RoMS) criterion (Rose & Hintz 2007). Based on those results, our selected stars were confirmed as δ Scuti variable stars. We determined the ephemerides of our sample by calculating the period of star oscillations and discussing light curve structure and pulsation behavior. The physical and geometrical parameters were also obtained. We collected a new dataset based on observed δ Scuti stars by the *Kepler* mission, TESS mission, and ASAS-SN catalog of variable stars. Then, we were able to deduce new $P-L$ relations for δ Scuti stars according to our results.

## 2. Observations and data reduction

The Kourovka Planet Search project (KPS) carried out photometric observations during the interval 2012–2015. This project utilized the robotic MASTER-II-Ural and Rowe-Ackermann Schmidt Astrograph (RASA) telescopes to search for new hot Jupiter exoplanets transiting their host stars. The MASTER-II-Ural telescope is operated at the Kourovka Astronomical Observatory of the Ural Federal University, Russia (57° N, 59° E), with a pair of Hamilton catadioptric tubes and a focal length of 400 mm corresponding to an Apogee Alta U16M CCD yielding an image scale of 1.8 arcsec pixel$^{-1}$. The second telescope, the RASA telescope, is a 279 mm Celestron CGEM mounted with a focal length of 620 mm, installed at the private observatory called Acton Sky Portal, Massachusetts, USA (43° N, 71° W). This telescope is equipped with an SBIG ST-8300M CCD camera which processes an image scale of 1.8 arcsec pixel$^{-1}$.

In this project, to investigate transiting exoplanets in a photometric survey, three different fields were selected. The CCD Images for the first field observations (named TF1) were collected with the robotic MASTER-II-Ural telescope from May to August 2012, with 50 s exposure times in the $R$ filter and 120 s exposure times in the $B$ and $V$ filters. The second field observations (named TF2) were obtained in 2013 and 2014 using the MASTER-II-Ural telescope with 50 s exposure times in the $R$ and $V$ filters, and 120 s exposure times in the $B$ and $I$ filters for determination of the color indices of the stars. The second telescope was involved in September 2014 with 50 s exposure times in the $R$ and 50 s exposure times in the $V$ filters. The third field (named TF3) was observed only in the $R$ filter with 50 s exposure times by the RASA telescope from January to April 2015. The MASTER-II-Ural telescope was also used in $BVRI$ in order to determine the color indices of the stars. Table 1 summarizes the obtained information from the photometric CCD data sets.

**Table 1.** The photometric observation parameters obtained from the KPS project.

| Field | Year | Telescope | Filter | Images | Hours |
|-------|------|-----------|--------|--------|-------|
| TF1 | 2012 | MASTER-II-Ural | $R$ | 3600 | 90 |
| TF1 | 2012 | MASTER-II-Ural | $BV$ | 1000 | |
| TF2 | 2013-2014 | MASTER-II-Ural | $VR$ | 4400 | 100 |
| TF2 | 2014 | MASTER-II-Ural | $BI$ | 500 | |
| TF2 | 2014 | RASA | $R$ | 8000 | 130 |
| TF2 | 2014 | RASA | $V$ | 485 | |
| TF3 | 2015 | RASA | $R$ | 7000 | 115 |
| TF3 | 2015 | MASTER-II-Ural | $BVRI$ | 50 | |

During the observation of three celestial fields, several variable stars were monitored. Their coordinates, magnitudes, and Gaia EDR3 distances are tabulated in Table 2. The variability of these stars was discussed by Burdanov et al. (2016).

The Image Reduction and Analysis Facility (IRAF) package (Tody 1986) was performed for the reduction of derived photometric data. In order to conduct aperture photometry, the PHOT task in the IRAF package was carried out for each frame according to the 2MASS Point Source Catalog[1] (Cutri et al. 2003). The size of the aperture taken for a particular frame was about 0.8×FWHM for the data from the MASTER-II-Ural telescope and 0.7×FWHM for the RASA telescope. The stars' brightness variations were corrected by the ASTROKIT program (Burdanov et al. 2014) in order to take into account the variations in atmospheric transparency.

**Table 2.** Coordinates, magnitudes, and Gaia EDR3[2] distances of the δ Scuti stars.

| Star | RA. (J2000) | Dec. (J2000) | $G$ (mag.) | $d(pc)$ |
|---|---|---|---|---|
| 2MASS 20250468+5026580 | 306.26949064028054 | 50.44941023165339 | 14.191 | 2836(105) |
| 2MASS 20262340+5005365 | 306.59755355716600 | 50.09346651224345 | 12.971 | 1733(38) |
| 2MASS 20274366+4944360 | 306.93190620757420 | 49.74329553821871 | 11.544 | 908(19) |
| 2MASS 20274367+5021300 | 306.93200854255696 | 50.35837570503766 | 12.561 | 1309(18) |
| 2MASS 20274485+5025395 | 306.93691416417610 | 50.42761943714574 | 10.926 | 478(3) |
| 2MASS 20274663+5121461 | 306.94430767928200 | 51.36281039402532 | 12.978 | 1827(43) |
| 2MASS 20274915+4935599 | 306.95478068368845 | 49.60001073955909 | 12.259 | 884(8) |
| 2MASS 20284384+5031252 | 307.18273675289225 | 50.52364421107788 | 12.938 | 2200(58) |
| 2MASS 20291369+5043247 | 307.30704732384066 | 50.72351787757314 | 12.770 | 1589(35) |
| 2MASS 20291725+4943570 | 307.32185591546260 | 49.73248990150829 | 12.964 | 1517(35) |
| 2MASS 20292279+5018015 | 307.34505980847100 | 50.30039489034873 | 14.065 | 1166(17) |
| 2MASS 20294536+5032540 | 307.43901478656835 | 50.54833727953713 | 11.062 | |
| 2MASS 20294695+4930547 | 307.44550661452183 | 49.51522665639457 | 12.454 | 996(145) |
| 2MASS 20295420+5032315 | 307.47594862190647 | 50.54202204195414 | 12.953 | 1410(43) |
| 2MASS 20302031+4943117 | 307.58463070187650 | 49.71994700960984 | 12.160 | 998(10) |
| 2MASS 20304189+4957269 | 307.67454527223900 | 49.95748835606642 | 11.851 | 884(7) |
| 2MASS 20304930+5104595 | 307.70536513818877 | 51.08315638333631 | 11.278 | 512(6) |
| 2MASS 20310164+5014147 | 307.75688217576120 | 50.23739249923349 | 11.924 | 1300(17) |
| 2MASS 20311156+5111105 | 307.79817639389460 | 51.18627232451892 | 11.397 | 1101(18) |
| 2MASS 20311900+4946378 | 307.82920178833340 | 49.77717970588341 | 11.705 | 901(8) |
| 2MASS 20320758+5044470 | 308.03160145014033 | 50.74641393343131 | 11.889 | 923(10) |
| 2MASS 20324010+5036368 | 308.16705801412210 | 50.61027363894143 | 11.699 | 619(5) |
| 2MASS 20325225+5054269 | 308.21770980874123 | 50.90745849209710 | 12.800 | 1185(17) |
| 2MASS 20341630+5043362 | 308.56791870926804 | 50.72670818271319 | 11.470 | 1101(20) |
| 2MASS 20341779+5041368 | 308.57411182701430 | 50.69357664258614 | 11.131 | 702(21) |
| 2MASS 20343224+4945589 | 308.63438643747770 | 49.76634925525544 | 10.065 | 657(6) |
| 2MASS 20344974+4953155 | 308.70725694735810 | 49.88764671616395 | 12.367 | 1328(16) |

### 3. Ephemerides calculation

Calculating the ephemerides of δ Scuti stars is the usual method for finding period changes. This shows how times of maxima may change over time in comparison to the reference ephemeris. We first calculated the period of the stars' oscillations using Period04 software (Lenz & Breger 2004). Period04 is a computer program dedicated to extracting the individual frequencies from the multiperiodic content of time series using Fourier analysis. After that, we extracted all light curves from the $BVR$ filters separately. Then, we determined the times of maxima ($T_0$) by fitting a sinusoidal curve to each light curve using the period that was calculated by Period04 analysis[3]. Because some of the data from the first observation night was insufficient, we chose $T_0$ based on observation nights with more complete and reliable data. Figure 1 shows how to select $T_0$ among some other visibility times of the maximum for 2MASS 20320758+5044470 during an observation night.

---

[1] The Two Micron All Sky Survey Point Source Catalog
[2] https://gea.esac.esa.int/archive/
[3] A supplement to this paper provides a machine readable table that contains times of maximum and observed filters for 27 stars.

All selected times of maximum were checked using the following equation and determined to be within the acceptable range,

$$T_{max} = T_0 + E \times P \quad (1)$$

Then, we plotted the time of maxima diagram in terms of epochs and then calculated the ephemeris for each. For this plot, we applied the MCMC approach (100 walkers, 10000 step numbers, and 200 burn-in) using the emcee package in Python (Foreman-Mackey et al. 2013). The calculated ephemeris for each star is shown in Table 3.

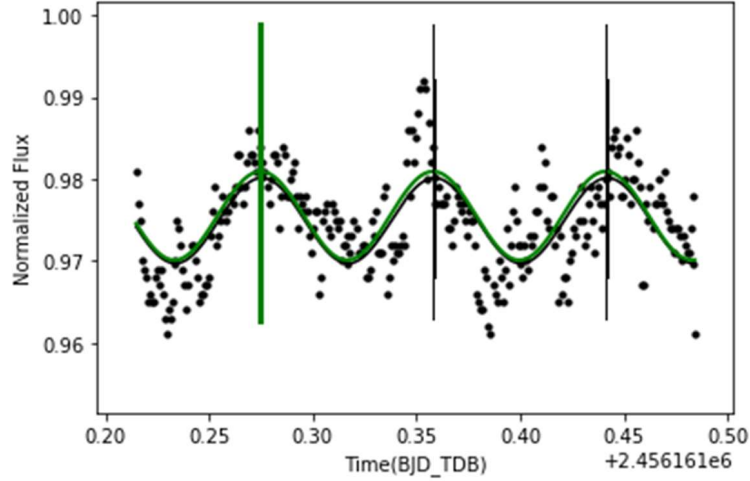

**Figure 1.** The black lines indicate the specified maximum times, whereas the green line shows the selected $T_0$. The green curve fitted on a light curve using calculated period for 2MASS 20320758+5044470 is most consistent with $T_0$=2456161.274335 (BJD$_{TDB}$).

**Table 3.** The calculated ephemeris for 27 stars. The maximum times are in BJD$_{TDB}$.

| Star | Ephemeris |
| --- | --- |
| 2MASS 20250468+5026580 | 2456270.309484(846) + 0.06543177(37) × $E$ |
| 2MASS 20262340+5005365 | 2456272.131292(883) + 0.13273525(83) × $E$ |
| 2MASS 20274366+4944360 | 2456140.310155(413) + 0.04184466(12) × $E$ |
| 2MASS 20274367+5021300 | 2456272.130515(329) + 0.04758513(11) × $E$ |
| 2MASS 20274485+5025395 | 2456366.529412(844) + 0.05166587(24) × $E$ |
| 2MASS 20274663+5121461 | 2456169.227651(873) + 0.07053278(35) × $E$ |
| 2MASS 20274915+4935599 | 2456270.375081(646) + 0.08018484(36) × $E$ |
| 2MASS 20284384+5031252 | 2456160.432461(1300) + 0.21939739(165) × $E$ |
| 2MASS 20291369+5043247 | 2456135.396293(1286) + 0.12570476(68) × $E$ |
| 2MASS 20291725+4943570 | 2456148.289354(1305) + 0.13092958(73) × $E$ |
| 2MASS 20292279+5018015 | 2456162.300346(1632) + 0.17006484(179) × $E$ |
| 2MASS 20294536+5032540 | 2456131.346189(824) + 0.09653837(63) × $E$ |
| 2MASS 20294695+4930547 | 2456412.435277(771) + 0.06710924(27) × $E$ |
| 2MASS 20295420+5032315 | 2456167.442096(1255) + 0.19545776(141) × $E$ |
| 2MASS 20302031+4943117 | 2456441.351810(796) + 0.05355508(19) × $E$ |
| 2MASS 20304189+4957269 | 2456134.349865(757) + 0.07786985(29) × $E$ |
| 2MASS 20304930+5104595 | 2456154.353676(648) + 0.05002574(20) × $E$ |
| 2MASS 20310164+5014147 | 2456400.408104(672) + 0.06989086(26) × $E$ |
| 2MASS 20311156+5111105 | 2456134.366705(401) + 0.06023030(23) × $E$ |
| 2MASS 20311900+4946378 | 2456496.336117(950) + 0.06807526(25) × $E$ |
| 2MASS 20320758+5044470 | 2456161.273686(657) + 0.08317405(30) × $E$ |
| 2MASS 20324010+5036368 | 2456162.282184(926) + 0.16782943(110) × $E$ |
| 2MASS 20325225+5054269 | 2456155.298711(752) + 0.06598773(24) × $E$ |
| 2MASS 20341630+5043362 | 2456358.532697(546) + 0.05115637(17) × $E$ |
| 2MASS 20341779+5041368 | 2456160.371278(572) + 0.10445878(39) × $E$ |
| 2MASS 20343224+4945589 | 2456124.333603(1900) + 0.13486517(185) × $E$ |
| 2MASS 20344974+4953155 | 2456169.228667(1177) + 0.08397630(81) × $E$ |

We plotted the phase-normalized flux curve for each star by integrating the data for all filters (Appendix B); and as a sample, Figure 2 shows the resulting curve of 2MASS 20250468+5026580.

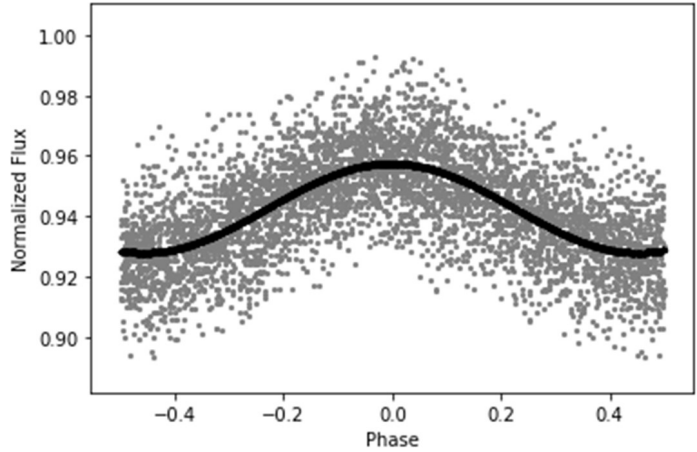

**Figure 2.** The phase-normalized flux curve for 2MASS 20250468+5026580.

We extracted the magnitudes of minimum and maximum from the light curves in all filters. The calculated values are given in Appendix A. The last column of the table is the mean of the apparent magnitude changes in all observed filters.

### 4. Calculating physical parameters

We extracted the physical parameters of the stars. First, distances of the stars were calculated using the Gaia EDR3 and then $M_v$ values were derived according to this study's average $V$ (mag). The extinction coefficient $A_v$ and its uncertainty were calculated utilizing the dust-maps Python package of Green et al. (2019) and then the interstellar reddening value, $E(B-V)$, was determined based on the following equation,

$$A_v = R_v \times E(B - V) \qquad (2)$$

where the coefficient $R_v = 3.1$ (Fitzpatrick 1999). The below equation was used to calculate $M_v$,

$$M_v = V - 5 \log d + 5 - A_v \qquad (3)$$

Using the relation $M_{bol} = M_V + BC$, the bolometric absolute magnitude was determined as well. In this relation, the bolometric correction ($BC$) was estimated based on the polynomial transformation equations derived by Flower (1996) that were corrected by Torres (2010).

The luminosity ($L$) can be calculated by using Pogson's relation (Pogson 1856),

$$L/L_\odot = 10^{((M_{bol_\odot} - M_{bol})/2.5)} \qquad (4)$$

The effective temperature ($T$) was taken from the TESS input catalog (Stassun et al. 2018). Depending on the $L$ and $T$ values, the radius ($R$) of the stars could be obtained from the relation,

$$R/R_\odot = \sqrt{\frac{L/L_\odot}{(T_{eff}/T_{eff\odot})^4}} \qquad (5)$$

Masses ($M$) of the examined δ Scuti stars (in solar masses $M_\odot$) can be calculated from the relation (Cox 1999),

$$\log M = 0.46 - 0.1 M_{bol} \qquad (6)$$

Then, the surface gravity was determined by

$$g = G_\odot (M/R^2) \qquad (7)$$

The uncertainties of the parameters (e.g., $M_v$, $L$, $M$, $R$, $\log g$) were calculated considering the error bars of the associated parameters such as errors in Gaia EDR3 distance, $V$ magnitude from observation, $A_v$, and TESS input catalog (TIC) temperature.

According to the calculations, the obtained global parameters of the δ Scuti stars were tabulated in Table 4. In the last column, the TESS $\log(g)$ values are given. Those values are comparable with this study's result.

**Table 4.** The physical parameters of the δ Scuti stars derived from this study.

| Star | $M_v$ (mag.) | $M_{bol}$ (mag.) | $L$ ($L_\odot$) | $T$ (K) | $R$ ($R_\odot$) | $M$ ($M_\odot$) | $\log(g)$ (cgs) | $\log(g)$ TESS |
|---|---|---|---|---|---|---|---|---|
| 2MASS 20250468+5026580 | 0.971(30) | 1.006(45) | 31.16(6) | 7548(123) | 3.27(17) | 2.29(26) | 3.77(23) | 3.65(9) |
| 2MASS 20262340+5005365 | 0.994(54) | 1.029(56) | 30.51(8) | 6797(184) | 3.99(14) | 2.28(33) | 3.59(20) | 3.51(9) |
| 2MASS 20274366+4944360 | 1.285(33) | 1.319(39) | 23.36(4) | 6896(178) | 3.39(18) | 2.13(20) | 3.71(25) | 3.66(10) |
| 2MASS 20274367+5021300 | 1.544(48) | 1.576(57) | 18.43(6) | 7479(163) | 2.56(17) | 2.01(30) | 3.92(25) | 3.89(9) |
| 2MASS 20274485+5025395 | 2.301(84) | 2.335(89) | 09.16(9) | 7432(209) | 1.83(16) | 1.68(30) | 4.14(24) | 4.18(8) |
| 2MASS 20274663+5121461 | 1.236(51) | 1.271(63) | 24.41(8) | 7385(183) | 3.02(18) | 2.15(29) | 3.81(23) | 3.72(8) |
| 2MASS 20274915+4935599 | 1.935(66) | 1.970(51) | 12.82(6) | 7045(166) | 2.41(14) | 1.83(19) | 3.94(22) | 3.82(9) |
| 2MASS 20284384+5031252 | 0.556(52) | 0.591(54) | 45.67(6) | 7779(107) | 3.71(18) | 2.52(34) | 3.70(26) | 3.44(12) |
| 2MASS 20291369+5043247 | 1.093(36) | 1.127(48) | 27.87(6) | 7597(364) | 3.05(18) | 2.22(20) | 3.82(20) | 3.68(10) |
| 2MASS 20291725+4943570 | 1.238(42) | 1.273(57) | 24.37(6) | 7056(105) | 3.31(18) | 2.15(27) | 3.73(21) | |
| 2MASS 20292279+5018015 | 3.553(91) | 3.587(93) | 02.89(9) | 6383(21) | 1.39(17) | 1.26(24) | 4.25(18) | 4.19(9) |
| 2MASS 20294695+4930547 | 0.689(57) | 0.724(59) | 40.40(6) | 7481(145) | 3.79(16) | 2.44(32) | 3.67(23) | |
| 2MASS 20295420+5032315 | 1.514(33) | 1.548(36) | 18.91(4) | 7111(121) | 2.87(18) | 2.02(17) | 3.83(18) | |
| 2MASS 20302031+4943117 | 1.610(51) | 1.642(57) | 17.35(6) | 7332(204) | 2.59(16) | 1.98(30) | 3.91(24) | 3.79(9) |
| 2MASS 20304189+4957269 | 1.569(39) | 1.602(45) | 18.00(5) | 7236(109) | 2.71(16) | 1.99(16) | 3.87(22) | 3.77(8) |
| 2MASS 20304930+5104595 | 2.494(96) | 2.528(99) | 07.67(9) | 7279(148) | 1.75(14) | 1.61(35) | 4.16(24) | 4.17(8) |
| 2MASS 20310164+5014147 | 0.929(38) | 0.964(42) | 32.39(5) | 7015(173) | 3.92(18) | 2.31(22) | 3.61(23) | 3.46(9) |
| 2MASS 20311156+5111105 | 0.946(30) | 0.981(45) | 31.89(4) | 7589(146) | 3.33(18) | 2.30(23) | 3.75(23) | 3.64(9) |
| 2MASS 20311900+4946378 | 1.472(97) | 1.505(98) | 19.68(9) | 6687(199) | 3.31(16) | 2.04(46) | 3.71(23) | 3.52(9) |
| 2MASS 20320758+5044470 | 1.866(41) | 1.901(46) | 13.66(5) | 7014(135) | 2.51(18) | 1.86(18) | 3.91(22) | 3.76(9) |
| 2MASS 20324010+5036368 | 2.635(80) | 2.670(84) | 06.73(9) | 7004(148) | 1.77(15) | 1.56(30) | 4.13(19) | 4.10(8) |
| 2MASS 20325225+5054269 | 2.248(90) | 2.283(98) | 09.61(9) | 7127(144) | 2.04(16) | 1.70(33) | 4.05(23) | 4.05(9) |
| 2MASS 20341630+5043362 | 0.841(31) | 0.875(33) | 35.16(4) | 8200(364) | 2.94(18) | 2.36(20) | 3.87(24) | 3.90(8) |
| 2MASS 20341779+5041368 | 1.638(69) | 1.673(72) | 16.86(8) | 6863(138) | 2.91(16) | 1.96(30) | 3.80(21) | 3.76(9) |
| 2MASS 20343224+4945589 | 0.293(63) | 0.328(72) | 58.18(8) | 8455(172) | 3.56(18) | 2.67(40) | 3.76(20) | 3.63(7) |
| 2MASS 20344974+4953155 | 0.921(93) | 0.954(97) | 32.69(9) | 7247(409) | 3.63(16) | 2.32(56) | 3.68(29) | 3.40(8) |

It should be noted that star 2MASS 20294536+5032540 did not have a Gaia parallax, so it was not possible to calculate its physical parameters.

## 5. Pulsation modes

Two main types of pulsation modes are pressure ($p$) and gravity ($g$) modes. For the $p$-mode pressure works as a restoring force and is made by acoustic waves with vertical motions for a star that has lost its equilibrium. For the $g$-mode, buoyancy (gravity) acts as a restoring force but gas motions are horizontal (Joshi & Joshi 2015). The pulsation modes are distinguished in terms of quantum numbers $(n, l, m)$, which describe the star's geometry. If the star's spherical symmetry is maintained due to its pulsation, the mode of pulsation is known as radial. Otherwise, the pulsation is non-radial. δ Scuti stars have periods between 0.02 and 0.25 days (McNamara et al. 2000). These stars have pulsations of radial and/or non-radial low-order $p$-mode (Joshi & Joshi 2015). The non-radial pulsations that have been found are low-overtones ($n = 0$ to 7) and low degree ($l \ll 3$) $p$-modes for δ Scuti stars.

The dominant frequency of pulsation for the stars in our sample was determined by the light curve analysis and using Period04 software. To identify the pulsation mode, at first we checked the periodogram of these stars. Figure 3 shows the periodogram of 2MASS 20250468+5026580 as an example, which was observed during 2012-2013 albeit not continuously. Therefore, the aliasing effect occurred in the observational data. The periodograms of the other stars can be found in Appendix B. There is a significant amount of noise and aliases in those periodograms. According to the fact that the data for some stars may not be accurate enough, and given that the stars were not observed continuously, we may not have been able to detect non-radial modes. However, calculating the pulsation constant, $Q$, can be useful to identify the $p$-mode order for our stars.

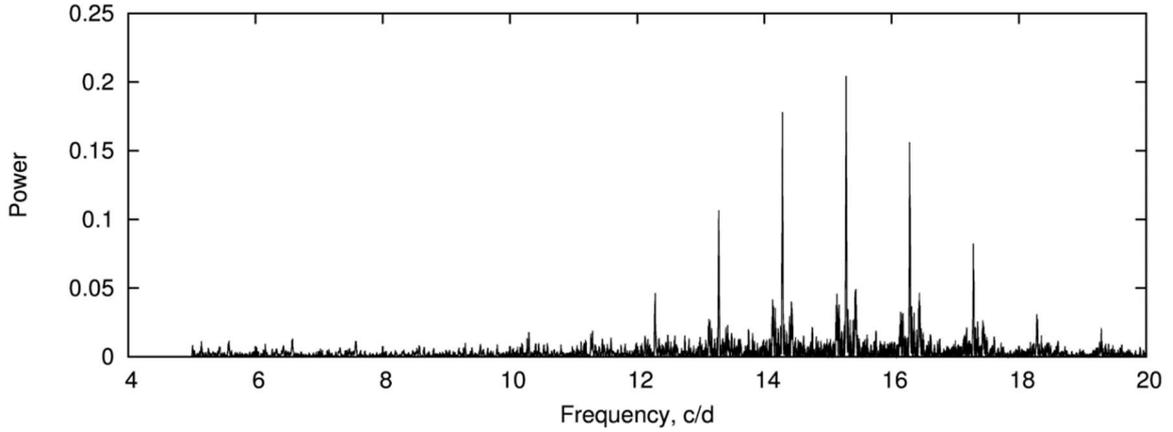

**Figure 3.** Periodogram of 2MASS 20250468+5026580 during the 2012-2013 observations.

Since the period is related to star structure through pulsation, the period is therefore related to the radius change but more accurately to the mean density if we assume that the mass of a star is constant. This is defined by the period-density relation,

$$Q = P\sqrt{\frac{\bar{\rho}}{\bar{\rho}_\odot}} \qquad (8)$$

where $\bar{\rho}$ and $\bar{\rho}_\odot$ are the mean densities of the target star and the Sun, respectively, and $P$ is the pulsation period. The equation can be written in terms of other parameters as Breger (1990) derived,

$$log\left(\frac{Q}{P}\right) = 0.1(M_{bol} - M_{bol\odot}) + 0.5\, log\frac{g}{g_{Sun}} + log\left(\frac{T_{eff}}{T_{eff\odot}}\right) \qquad (9)$$

Using the equation, we calculated $Q$ values for the 26 δ Scuti stars listed in Table 5. For this calculation, we considered the period ($P$) from Table 3 and other parameters $M_{bol}$, $lo\,g(g)$, and $T_{eff}$ from Table 4. We have determined the $p$-mode order for each pulsating star. According to the second table of Breger (1979) and our calculated $Q$ values, it was found that $p$-mode is fundamental for 11 stars, the first overtone for 4 stars, the second overtone for 3 stars, and the third overtone for 5 stars. For the remaining 3 stars $Q$ values are not in the range. Referring to the distribution of $Q$ values for a sample of 75 δ Scuti stars shown in figure seven of North et al. (1997), it is determined that the $p$-mode of 11 stars is fundamental, for 4 stars it is the first overtone, for 3 stars the second overtone, and for 5 stars it is the third overtone. However, for 3 other stars their $Q$ values are not in the distribution. The amount of uncertainty for $Q$ values is calculated according to the amount of uncertainty in the parameters involved.

**Table 5.** Estimated $Q$ values for 26 δ Scuti stars. In the table, "F" refers to the fundamental mode, "1" refers to the first overtone, "2" refers to the second overtone and "3" refers to the third overtone.

| Star | $Q$(d) | $p$-mode (Breger 1979) | $p$-mode (North et al. 1997) |
| --- | --- | --- | --- |
| 2MASS 20250468+5026580 | 0.017(5) | 2 | 2 |
| 2MASS 20262340+5005365 | 0.025(7) | F | F |
| 2MASS 20274366+4944360 | 0.010(3) | 3 | 3 |
| 2MASS 20274367+5021300 | 0.016(5) | 2 | 2 |
| 2MASS 20274485+5025395 | 0.027(9) | F | F |
| 2MASS 20274663+5121461 | 0.020(6) | 1 | 1 |
| 2MASS 20274915+4935599 | 0.029(9) | F | F |
| 2MASS 20284384+5031252 | 0.049(18) | F | F |
| 2MASS 20291369+5043247 | 0.035(11) | F | F |
| 2MASS 20291725+4943570 | 0.032(9) | F | F |
| 2MASS 20292279+5018015 | 0.116(29) | Not in range | Not in range |
| 2MASS 20294695+4930547 | 0.014(4) | 3 | 3 |
| 2MASS 20295420+5032315 | 0.057(14) | Not in range | Not in range |
| 2MASS 20302031+4943117 | 0.018(6) | 1 | 1 |
| 2MASS 20304189+4957269 | 0.025(7) | F | F |
| 2MASS 20304930+5104595 | 0.028(9) | F | F |
| 2MASS 20310164+5014147 | 0.014(4) | 3 | 3 |
| 2MASS 20311156+5111105 | 0.015(4) | 3 | 3 |
| 2MASS 20311900+4946378 | 0.016(5) | 2 | 2 |
| 2MASS 20320758+5044470 | 0.029(8) | F | F |
| 2MASS 20324010+5036368 | 0.089(25) | Not in range | Not in range |
| 2MASS 20325225+5054269 | 0.030(9) | F | F |
| 2MASS 20341630+5043362 | 0.016(5) | 3 | 3 |
| 2MASS 20341779+5041368 | 0.029(9) | F | F |
| 2MASS 20343224+4945589 | 0.033(9) | 1 | 1 |
| 2MASS 20344974+4953155 | 0.018(8) | 1 | 1 |

According to Table 5, in the two models, three stars are not in the range. We could not estimate $T_{eff}$ and $log(g)$ very well because the stars do not have spectroscopic data. Additionally, if any of these stars are a binary system this affects $T_{eff}$, $log(g)$, and $M_{bol}$ calculations that will change the $Q$ value.

## 6. Period-Luminosity relation
### 6.1. Previous studies and their results
It has been found that some groups of pulsating stars show a $P - L$ relation, such as the delta Cepheid and RR Lyrae stars. For pulsating stars, the period of the fundamental mode of oscillation is related to the mean density:

$$P \sim (G\bar{\rho})^{-1/2} \propto \left(\frac{M}{R^3}\right)^{-1/2} \qquad (10)$$

Since $L \propto R^3$, we have that

$$P \propto L^{3/4} M^{-1/2} \quad (11)$$

which implies the existence of a period-luminosity relation.

The first $P - L$ relation for δ Scuti stars was commonly determined by referring to the fact that the stars obey the same $P - L$ relation as Cepheids and can be used as standard candles. Fernie (1992) considered a combination of 28 Cepheids and 28 fundamental-mode δ Scuti stars to improve the $P - L$ relation. By combining $M_v$ values of the δ Scuti stars obtained by uvbyβ data with $M_v$ values of Cepheids and a fit to $M_v$ values, it was shown that δ Scuti stars obey the same $P - L$ relation as Cepheids. A few years later, with the aid of the Baade-Wesselink method used by Laney and Stobie (1995) and Hipparcos parallaxes, Laney et al. (2002) obtained a $P - L$ relation for both Cepheid and HADS stars. In the investigation of this new relation, several galactic δ Scuti stars were considered by McNamara et al. (2004), and the Hipparcos parallaxes were used to drive two equations of $P - L$ relation as a function of metallicity for variable stars. After these results, McNamara kept improving the $P - L$ relation with a fit to the $M_v$ values of δ Scuti stars calculated by Hipparcos parallaxes and $M_v$ values of Cepheids by other techniques (McNamara al. 2007). Following up the finding of a new $P - L$ relation, McNamara (2011) presented the latest relation by a fit to $M_v$ values of HADS stars. Distances of four galaxies and two globular clusters were derived using the latest equation by McNamara.

In terms of a recent empirical study on the $P - L$ relation of δ Scuti stars, we should mention Ziaali et al. (2019). With the aid of Gaia DR2 parallaxes and extinction corrections, they calculated $M_v$ values of 1352 δ Scuti stars, where 1124 of them were observed by *Kepler* during its four-year mission and the remaining stars were cataloged by Rodríguez et al. (2000). They plotted the $M_v$ values against the $\log P$ of stars after restricting *Kepler* stars to 601 samples that have pulsation of semi-amplitudes above one millimagnitude to make a better comparison between ground-based samples and samples obtained from space observation. Their inferred $P - L$ relation is close to the relation obtained by McNamara (2011). The results of the Ziaali et al. (2019) study indicate that many stars fall along a ridge close to the published $P - L$ relation of the radial fundamental mode in δ Scuti stars. Also, an excess of stars in a second ridge corresponds to stars having a dominant period that is half that of the main ridge. A significant number of δ Scuti stars from the ASAS-SN catalog was used by Jayasinghe et al. (2020) to improve the $P - L$ relation although the equation is presented without uncertainty. They used Gaia DR2 parallaxes to calculate the distance and then $M_v$ of the stars. They presented two $P - L$ relations for both the fundamental mode and the average overtone mode. Table 6 shows all the studied $P - L$ relations mentioned above for the fundamental mode.

**Table 6.** $P - L$ relations obtained for the fundamental mode of δ Scuti stars.

| Relation | Reference |
|---|---|
| $M_v = (-2.92 \pm 0.030) \log P - (1.203 \pm 0.029)$ | (Fernie 1992) |
| $M_v = (-2.92 \pm 0.004) \log P - (1.29 \pm 0.04)$ | (Laney et al. 2002) |
| $M_v = -2.90 \log P - 0.190[Fe/H] - 1.26 \quad [Fe/H] \gg -1.5$ <br> $M_v = -2.90 \log P - 0.089[Fe/H] - 1.11 \quad [Fe/H] < -1.5$ | (McNamara et al. 2004) |
| $M_v = (-2.90 \pm 0.05) \log P - (1.27 \pm 0.05)$ | (McNamara et al. 2007) |
| $M_v = (-2.89 \pm 0.13) \log P - (1.31 \pm 0.10)$ | (McNamara 2011) |
| $M_v = (-2.96 \pm 0.06) \log P - (1.36 \pm 0.06)$ | (Ziaali et al. 2019) |
| $M_v = -3.223 \log(P/0.1d) + 1.438$ | (Jayasinghe et al. 2020) |

### 6.2. New $P - $L relation for δ Scuti stars

Space telescopes have produced continuous, long-term light curves of a vast number of variable stars. For instance, a unique data set was obtained with the TESS mission to study main-sequence A and F type stars.

Furthermore, the *Kepler* mission published a huge space database, and the high precision *Kepler* light curves made it ideal for discovering new pulsating variable stars. Uytterhoeven et al. (2011) have studied and classified more than 750 *Kepler* samples of $\gamma$ Dor and δ Scuti candidates.

In order to present a new $P - L$ relation for δ Scuti stars, we have used 196 stars studied by Uytterhoeven et al. (2011) and six samples from Bradley et al. (2015). Our database is also composed of 11 TESS samples identified as δ Scuti stars studied by Antoci et al. (2019). We used 281 δ Scuti candidates from the ASAS-SN catalog. In addition to the samples described, we have used a sample of our 26 stars. We calculated the absolute magnitude, $M_v$, for each star using equation 3 of this study. We obtained apparent magnitude, $V$, from the APASS9, ASAS-SN, All-sky Compiled, The PASTEL, Tycho Input, UCAC2, UCAC4, and FON Astrographic catalogs. Using the Gaia EDR3 parallax[4] and extinction correction, the distances to all sample stars were calculated in parsecs. $A_v$, was obtained with the Bayestar 19 dust map of Green et al. (2019).

To understand this dataset characteristic, we used a Machine Learning "Support Vector Regression" to find the best fit to describe the whole dataset. Support Vector Machine (SVM) analysis is a common machine learning tool for classification and regression (Boser et al. 1992). In this study, we used SVM for regression with the epsilon parameter called Epsilon-Support Vector Regression. For using SVM in regression problems, a margin of tolerance (epsilon) is set as an approximation in the SVM model.

We used a global approach on all stars to significantly obtain the best linear regression to characterize all data clusters and found initial values for each model in dense areas before finding multi-fit lines to classify stars based on the $M_v$ and $P$ relationship. According to a 2D histogram, a heatmap is computed from a count of stars grouped by their two parameters ($M_v$ and $logP$) coordinates into bins. The darker color in each bin shows more stars in that bin. The many darker bins are close to each other indicating dense areas. These areas are the best places to look for a relationship between the data and classification.

To find the first-degree linear model initial parameter, we used the Epsilon-insensitive SVM (ε-SVM) regression based on our Python[5] code. SVM regression (SVR) uses different hyperparameters[6] to find the best fit in the dataset. The final model with a Mean Square Error (MSE) value of 0.177471 showed that -1.4[7] is the best coefficient in a first-degree equation to describe the dataset generally as a first-degree linear model initial parameter. This initial value, which was calculated at -1.4, is essential to finding the best model for each determined class of stars. The coefficient helps to limit the range of model samples to accrue band, so we can ensure that the model will find the best fit for each class of stellar target data.

This δ Scuti stars dataset contains two main star physical parameters, $M_v$ and $logP$. To determine a correlation between these two parameters based on data distribution, we used a 2D histogram (bin=20). There are some dense areas visible on the histogram. As a result, we were able to classify the data using machine learning. We expected stars in each dense region to have the same physical characteristics and parameter correlation based on the physical nature of the data. So, we needed to develop a multi-model for fitting each dense area separately.

To achieve this aim, we designed a hyperparameter linear model to find the best fit based on maximum star counts in each dense area with less than 0.01 error in the $M_v$ parameter prediction. For the initial parameter value, we used -1.4 as the best coefficient globally. To find the best fit in each dense field, the model tested over 100,000 parameters with different ranges based on the initial value.

---

[4] The Gaia DR3 will be based on 34 months of observation and has better accuracy than the Gaia DR2. All data in Gaia EDR3 is copied unchanged in Gaia DR3.
[5] Python 3.8, Scikit-learn 0.24
[6] kernel='poly', C=100, gamma='auto', epsilon=0.3, coef0=0.5
[7] $f(x) = -1.4x + b$

The result revealed that there are fits for each area separately to explain the parameter correlation between "$M_v$" and "$logP$" of stars in that specific area. We have recognized three additional areas with the best conditions in this model, which are identified as the areas of stars with an overtone mode of pulsation.

Figure 4 shows the *Kepler*, TESS, and ASAS-SN sample stars, as well as our 26 stars of various shapes. The black dashed line is the McNamara (2011) $P - L$ relation for the High-amplitude δ Scuti stars, the dashed blue line shows the $P - L$ relation derived by Ziaali et al. (2019), and the dashed purple line is related to the Jayasinghe et al. (2020) study. The fitted line to the dense area of the sample stars near the fundamental edge, which is shown by a solid green line in Figure 4, has the equation of

$$M_v = (-3.200 \pm 0.010)logP - (1.599 \pm 0.010) \qquad (12)$$

for the $P - L$ relation that is presented by this study for the fundamental mode δ Scuti stars. According to machine learning classification, the three areas identified as overtone mode areas, our 26 studied δ Scuti stars, and the $P - L$ relations for the overtone modes of δ Scuti stars are obtained. The following equations are related to the overtone area.

First overtone: $M_v = (-3.461 \pm 0.020)logP - (2.719 \pm 0.020)$ \qquad (13)
Second overtone: $M_v = (-2.885 \pm 0.030)logP - (2.445 \pm 0.030)$ \qquad (14)
Third overtone: $M_v = (-2.905 \pm 0.030)logP - (2.720 \pm 0.030)$ \qquad (15)

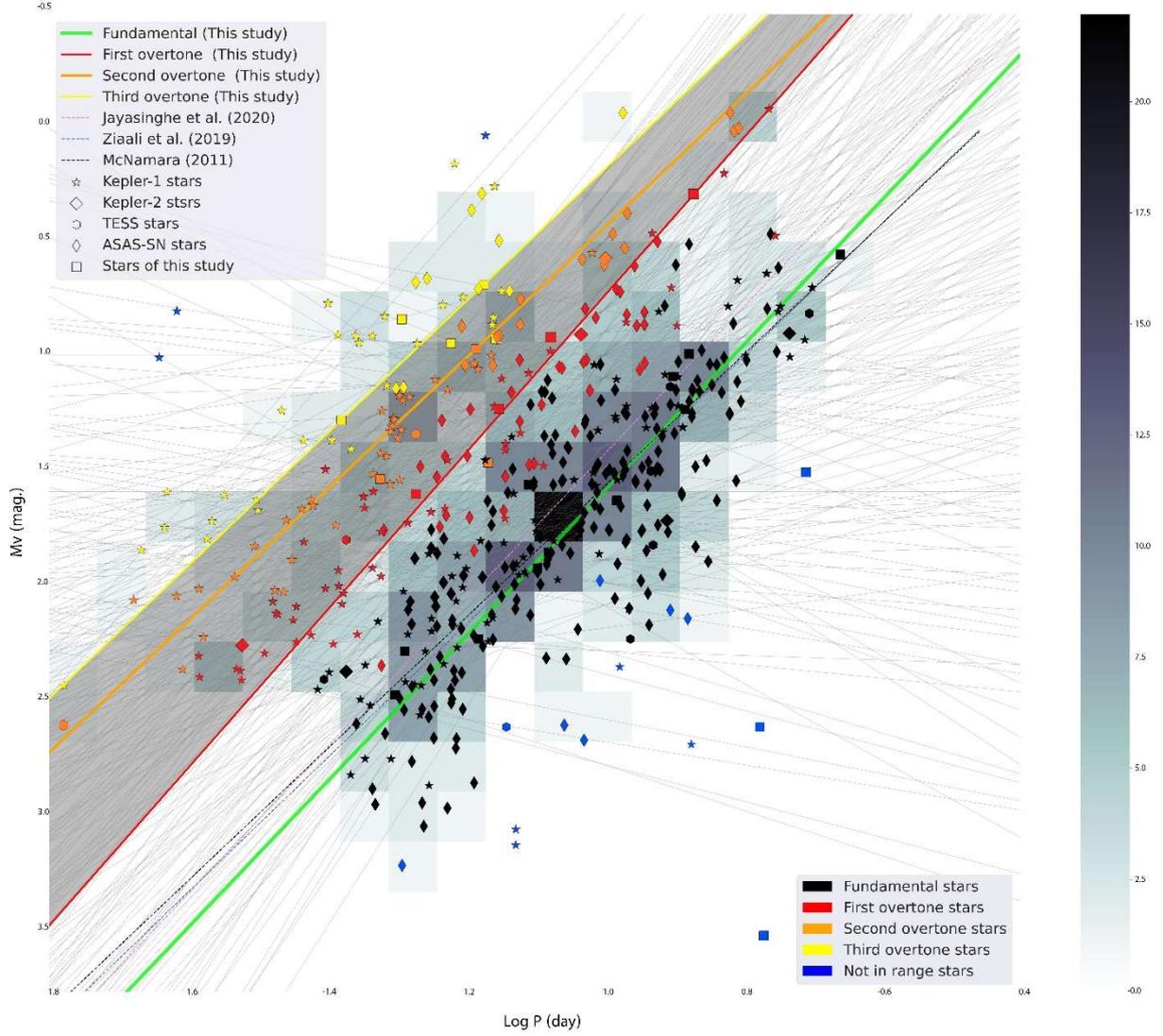

**Figure 4.** $P-L$ diagram of this study for δ Scuti stars. The solid green line represents our new derived $P-L$ relation to the fundamental mode. Also, the solid red line for the first overtone mode, the solid orange line for the second overtone, and the solid yellow line for the third overtone. The shaded area in the diagram shows the overtone area according to δ Scuti stars from our database.

## 7. Discussion of the results

We analyzed the multi-colored light curves of 27 δ Scuti stars obtained with photometric data observed between 2012 and 2015 in the KPS project.

To obtain the period and the ephemeris, we determined the times of maxima from the light curves and the initial epochs based on estimating the period of 27 stars by using Period04 software. We fitted the maxima times in terms of epochs to determine an ephemeris using a Python code based on the MCMC method.

We studied the light curves, pulsation modes, and the physical parameters of the stars. The absolute magnitude and physical parameters of the 26 δ Scuti stars were extracted using the results of the light curve analysis and Gaia EDR3 parallaxes. For star 2MASS 20294536+5032540, a Gaia EDR3 parallax does not exist; hence, we were not able to calculate the physical parameters and pulsation mode for this star.

We determined the $p$-mode order for stars, according to Breger (1979) and North et al. (1997), and our calculated $Q$ values. Based on the fact that the physical parameters for the stars and subsequent values are not obtained from spectroscopy, uncertainty is conceivable in the mode identification of our stars.

The TIC effective temperatures were used to compute physical parameters such as the radius and mass of the stars. It was found that the calculated $log(g)$ parameters (see Table 4) are in agreement with the TIC $log(g)$ values. Additionally, the pulsation constants of the stars were calculated by using the well-known equation of pulsation period and density.

The positions of the stars in the H-R diagram are shown in Figure 5. The evolutionary tracks of stars with masses between 1.5 and 2.5 solar masses are taken from the study of Kahraman Aliçavuş et al. (2016).
The star 2MASS 20292279+5018015 is not in the δ Scuti stars area, it is under the ZAMS, and its pulsation constant, Q, is not in the expected range (Table 5). There is a possibility that the parameters of this star are wrong, and can be commented on its variability in the future based on spectroscopic data. However, given the value of the star's luminosity derived from the Gaia EDR3 parallax, one can expect that the star has a different metallicity; therefore, the instability strip depending on metallicity can be changed. There is a similar condition for 2MASS 20324010+5036368 located close to the $\gamma$ Doradus instability strip. This star could be a $\gamma$ Doradus or δ Scuti - $\gamma$ Doradus hybrid star. According to the behavior of these stars, it could be a future study case.
Stars at the top left and out of the blue edge can be real Delta Scuti stars. After the *Kepler* mission, as pointed out by Uytterhoeven et al. (2011) some δ Scuti stars can be found outside of their instability strip. Additionally, one of these stars, 2MASS 20343224+4945589, according to its position in the H-R diagram and the frequency spectrum, could be a hot $\gamma$ Doradus star (Balona et al. 2015, Balona et al. 2016, Kahraman Aliçavuş et al. 2020).

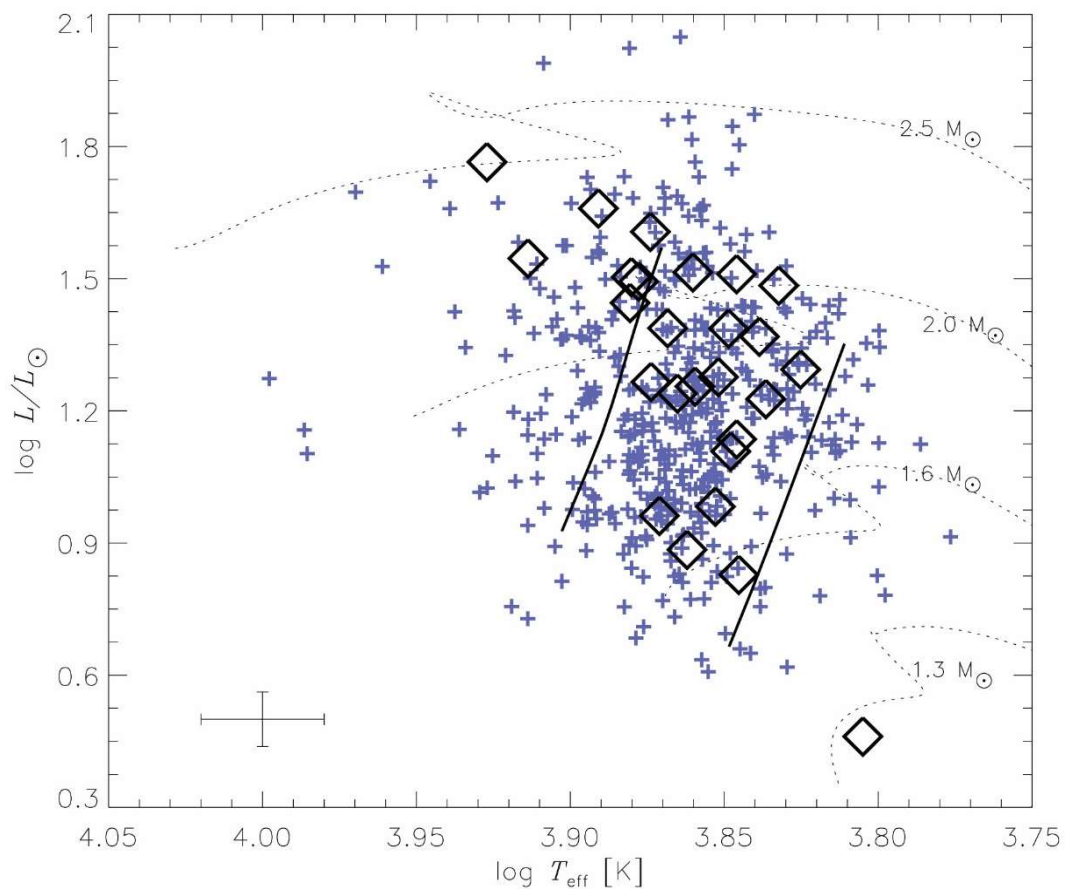

**Figure 5.** The instability strip of the δ Scuti stars (Dupret et al. 2005). This study's stars are shown in a rhombus shape. The average uncertainties of the parameters were estimated by a Monte Carlo process. The evolutionary tracks were taken from the study of Kahraman Aliçavuş et al. (2016).

Besides the presented photometric analysis of the 26 stars, we updated the δ Scuti period-luminosity relation using a number of δ Scuti stars and currently analyzed stars. A total of 520 stars were used. These stars are located on the H-R diagram between the blue and red edges of the Instability Strip (IS) and above the ZAMS, which is consistent with the main population of δ Scuti stars as discussed by Breger (1990). These borders are not absolute, but pulsation beyond these borders is less probable (Murphy et al. 2019).

We investigated the $P - L$ relation and improved measurement accuracy. Thus, we have used Gaia EDR3 data which increased parallax accuracy by 30% over Gaia DR2 data (Gaia Collaboration, Brown et al. 2020). The extinction correction for the sample stars is calculated using 3D dust maps that consider the distance module. To recognize and classify data, we have used machine learning classification and fitted lines to the fundamental and overtone modes areas. As a result, we present a new $P - L$ relation for the fundamental mode and overtone modes of δ Scuti variables.

**Acknowledgments**

This manuscript was prepared by cooperation between the International Occultation Timing Association Middle East section (IOTA/ME) and Tafresh University, Tafresh, Iran. This group activity happened during the Eighth Summer School of Astronomy, held between 21-26 August 2020.

This work has made use of data from the European Space Agency (ESA) mission Gaia (http://www.cosmos.esa.int/gaia), processed by the Gaia Data Processing and Analysis Consortium (DPAC, http://www.cosmos.esa.int/web/gaia/dpac/consortium). Funding for the DPAC has been provided by national institutions, particularly the institutions participating in the Gaia Multilateral Agreement. Also, we used the 2MASS catalog (https://irsa.ipac.caltech.edu/Missions/2mass.html), the TESS's results (https://mast.stsci.edu/), and the ASAS-SN catalog (https://asas-sn.osu.edu/variables).

"IRAF is distributed by the National Optical Astronomy Observatories, which are operated by the Association of Universities for Research in Astronomy, Inc., under cooperative agreement with the National Science Foundation". We would like to thank the Vienna Asteroseismology, and TOPS groups for improving on the Period04 software package.

This work was supported by the Ministry of Science and Education, FEUZ-2020-0030. The machine learning section of this study has been performed according to the scientific agreement with Raderon Lab Inc. (https://raderonlab.ca) with contract number R\AST\2021\1001. The authors would like to appreciate Dr. Fahri Alicavus and Dr. Somayeh Khakpash for their contributions to the research. Also, great thanks to Paul D. Maley for editing the text. The authors would like to thank the reviewer for comments and suggestions that helped to improve the paper.

**Appendix A**

**Table A1.** Minimum and maximum apparent magnitude and the difference between them for each star in $BVR$ filters.

| Star | $R_{\max}$ | $R_{\min}$ | $\Delta R$ | $V_{\max}$ | $V_{\min}$ | $\Delta V$ | $B_{\max}$ | $B_{\min}$ | $\Delta B$ | $\Delta(mag)$ |
|---|---|---|---|---|---|---|---|---|---|---|
| 2MASS 20250468+5026580 | 14.008 | 14.037 | 0.029 | 14.403 | 14.441 | 0.038 | | | | 0.033 |
| 2MASS 20262340+5005365 | 12.749 | 12.767 | 0.018 | 13.181 | 13.218 | 0.038 | | | | 0.028 |
| 2MASS 20274366+4944360 | 11.346 | 11.362 | 0.016 | 11.744 | 11.757 | 0.014 | | | | 0.015 |
| 2MASS 20274367+5021300 | 12.391 | 12.404 | 0.013 | 12.714 | 12.727 | 0.012 | 13.296 | 13.313 | 0.017 | 0.014 |
| 2MASS 20274485+5025395 | 10.794 | 10.801 | 0.006 | | | | 11.531 | 11.538 | 0.007 | 0.007 |
| 2MASS 20274663+5121461 | 12.822 | 12.835 | 0.014 | 13.128 | 13.145 | 0.018 | 13.689 | 13.714 | 0.024 | 0.019 |
| 2MASS 20274915+4935599 | 12.058 | 12.081 | 0.023 | 12.467 | 12.497 | 0.030 | 13.199 | 13.237 | 0.038 | 0.030 |
| 2MASS 20284384+5031252 | 12.761 | 12.779 | 0.018 | 13.137 | 13.158 | 0.020 | 13.761 | 13.788 | 0.027 | 0.022 |
| 2MASS 20291369+5043247 | 12.588 | 12.614 | 0.027 | 12.942 | 12.981 | 0.040 | 13.575 | 13.622 | 0.047 | 0.038 |
| 2MASS 20291725+4943570 | 12.740 | 12.772 | 0.032 | 13.200 | 13.241 | 0.041 | 13.994 | 14.045 | 0.051 | 0.041 |
| 2MASS 20292279+5018015 | 13.878 | 13.911 | 0.033 | 14.379 | 14.412 | 0.033 | | | | 0.033 |
| 2MASS 20294536+5032540 | 10.732 | 10.742 | 0.010 | | | | 11.570 | 11.583 | 0.012 | 0.011 |
| 2MASS 20294695+4930547 | 11.347 | 11.352 | 0.005 | | | | 12.381 | 12.389 | 0.007 | 0.006 |
| 2MASS 20295420+5032315 | 12.385 | 12.407 | 0.021 | 12.767 | 12.788 | 0.021 | 13.410 | 13.432 | 0.022 | 0.021 |
| 2MASS 20302031+4943117 | 11.970 | 11.976 | 0.005 | 12.341 | 12.348 | 0.007 | 13.028 | 13.038 | 0.010 | 0.007 |
| 2MASS 20304189+4957269 | 11.649 | 11.656 | 0.007 | 12.056 | 12.068 | 0.012 | 12.800 | 12.812 | 0.012 | 0.010 |
| 2MASS 20304930+5104595 | 11.127 | 11.136 | 0.009 | | | | 11.912 | 11.927 | 0.015 | 0.012 |
| 2MASS 20310164+5014147 | 11.736 | 11.743 | 0.006 | 12.126 | 12.134 | 0.008 | 12.816 | 12.826 | 0.010 | 0.008 |
| 2MASS 20311156+5111105 | 11.250 | 11.292 | 0.042 | 11.471 | 11.530 | 0.059 | 11.958 | 12.023 | 0.065 | 0.055 |
| 2MASS 20311900+4946378 | 11.493 | 11.522 | 0.029 | 11.917 | 11.953 | 0.036 | 12.663 | 12.709 | 0.046 | 0.037 |
| 2MASS 20320758+5044470 | 11.711 | 11.721 | 0.011 | 12.056 | 12.070 | 0.015 | 12.669 | 12.689 | 0.020 | 0.015 |
| 2MASS 20324010+5036368 | 11.529 | 11.537 | 0.008 | 11.861 | 11.871 | 0.011 | 12.441 | 12.452 | 0.011 | 0.010 |
| 2MASS 20325225+5054269 | 12.624 | 12.635 | 0.010 | 12.933 | 12.941 | 0.008 | 13.492 | 13.511 | 0.019 | 0.012 |
| 2MASS 20341630+5043362 | 11.357 | 11.361 | 0.005 | 11.565 | 11.570 | 0.005 | 11.996 | 12.003 | 0.007 | 0.006 |
| 2MASS 20341779+5041368 | 10.966 | 10.976 | 0.010 | 11.309 | 11.320 | 0.011 | 11.892 | 11.910 | 0.018 | 0.013 |
| 2MASS 20343224+4945589 | 9.917 | 9.934 | 0.018 | | | | 10.722 | 10.749 | 0.027 | 0.022 |
| 2MASS 20344974+4953155 | 12.138 | 12.158 | 0.020 | 12.441 | 12.461 | 0.020 | | | | 0.020 |

**Appendix B**

Periodograms and light curves of the stars were observed intermittently from 2012-2015.

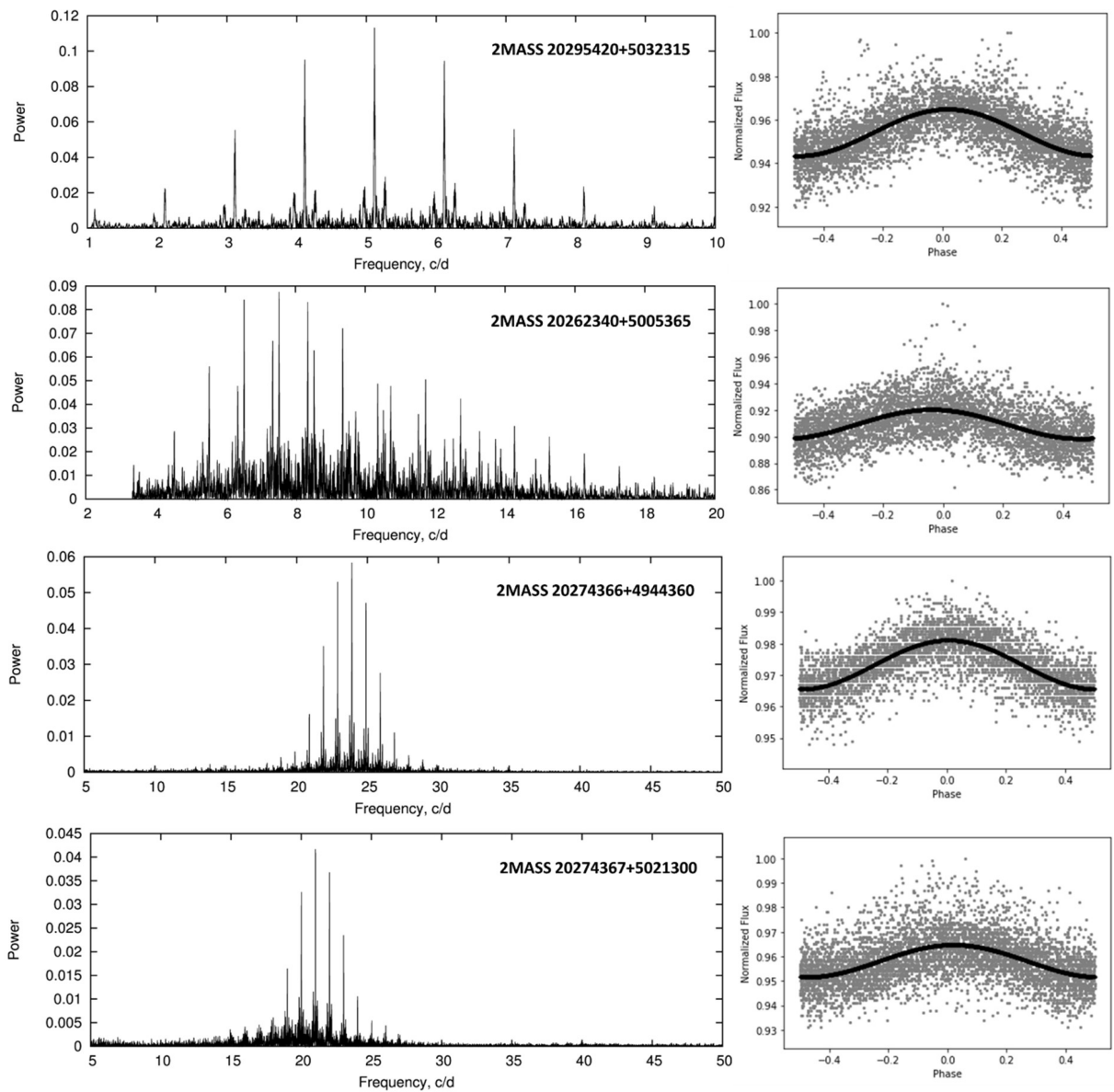

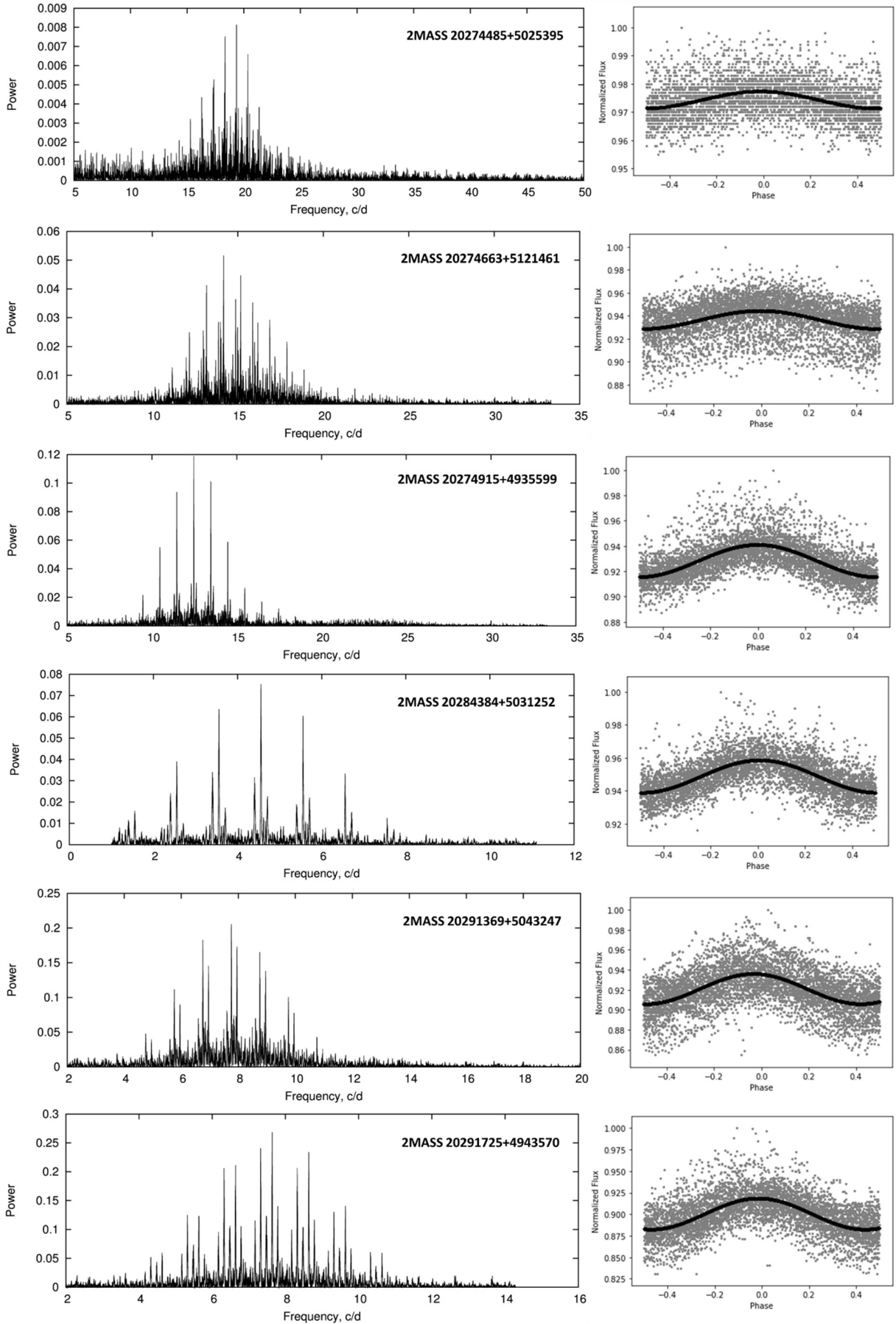

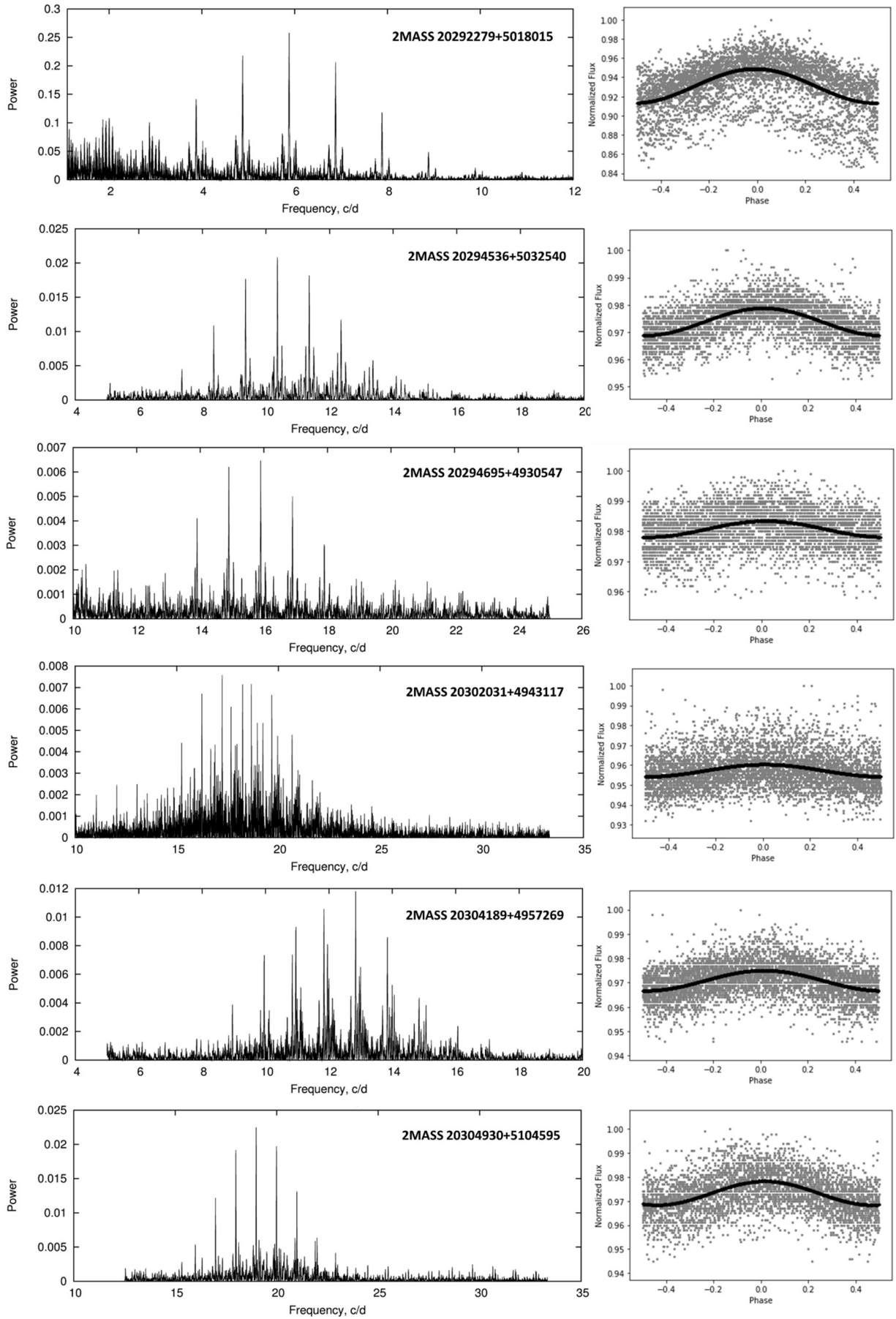

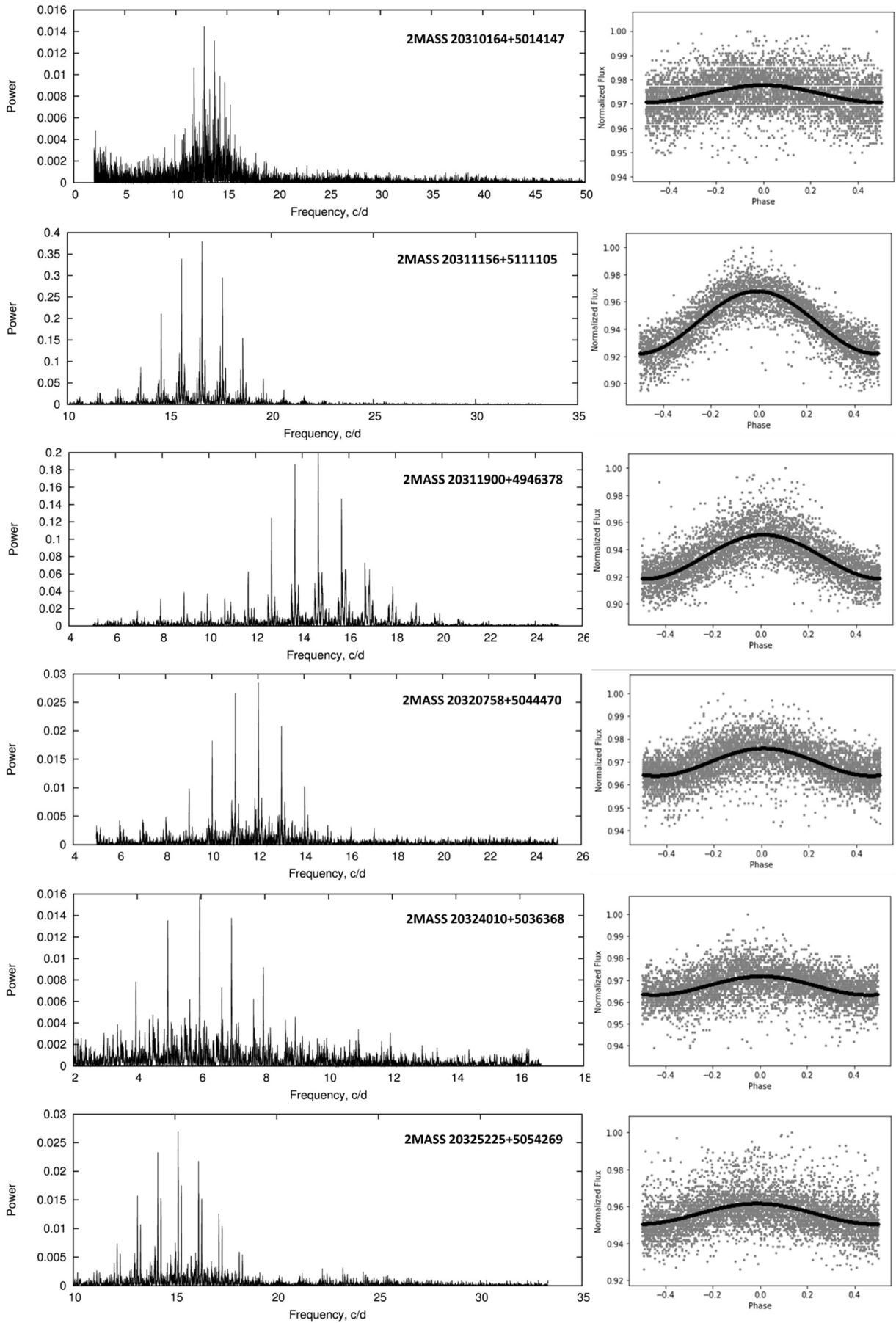

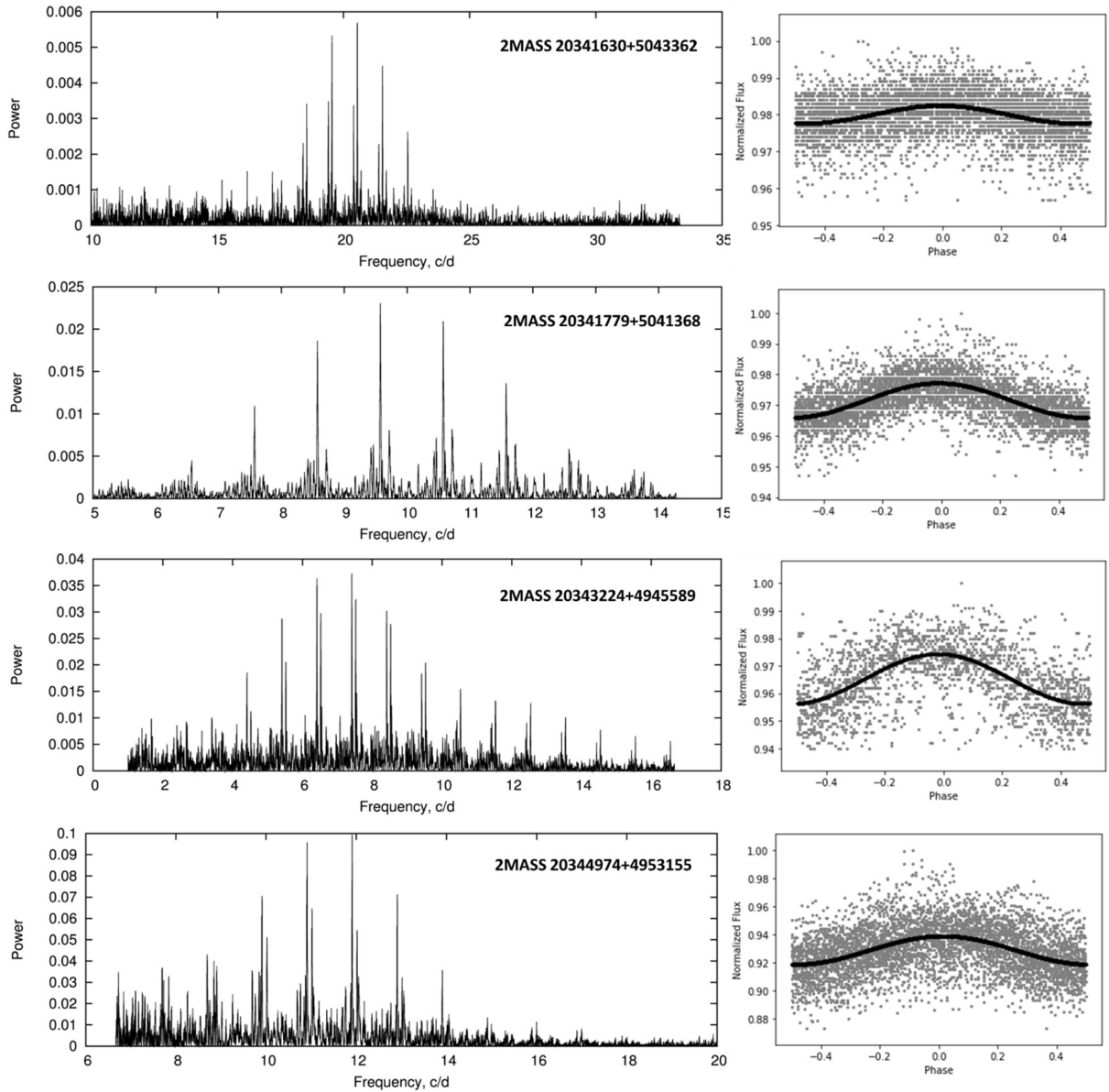